\documentclass{article}
\usepackage{arxiv}
\usepackage[utf8]{inputenc}
\usepackage[T1]{fontenc}

\usepackage[square,numbers,sort&compress]{natbib}
\usepackage{fontawesome5}
\usepackage{hyperref}
\usepackage{url}
\usepackage{booktabs}
\usepackage{amsfonts}
\usepackage{nicefrac}
\usepackage{microtype}
\usepackage{lipsum}
\usepackage{graphicx}
\usepackage{threeparttable}
\usepackage{doi}
\usepackage{multirow}
\usepackage{multicol}
\usepackage{xcolor}
\usepackage{subcaption}
\usepackage[normalem]{ulem}
\usepackage{amsmath,amsfonts,bm}
\usepackage{amssymb,amsthm}
\usepackage{enumitem}
\usepackage{adjustbox}
\usepackage{blindtext}
\usepackage{float}
\usepackage{array}
\usepackage{makecell}
\usepackage[perpage]{footmisc}
\usepackage{wrapfig}
\usepackage{tcolorbox}
\usepackage{CJKutf8}
\usepackage{multicol}

\usepackage{tcolorbox}
\tcbuselibrary{breakable, skins} 
\usepackage{CJKutf8}
\usepackage{tikz} 

\definecolor{casebg}{RGB}{236, 241, 245} 
\definecolor{thinkgray}{RGB}{80, 80, 80}

\newtcolorbox{CaseStudyBox}{
  colback=casebg,
  colframe=black,\textbf{}
  boxrule=1.2pt,
  arc=4mm,
  boxsep=8pt,
  left=10pt, right=10pt, top=10pt, bottom=10pt,
  breakable,
  enhanced,
  parbox=false,  
}

\usepackage{cleveref}

\setlist[itemize,1]{leftmargin=\dimexpr 18pt}
\setlist[enumerate,1]{leftmargin=\dimexpr 18pt}

\title{
\raisebox{-0.1\height}{\includegraphics[width=0.04\textwidth]{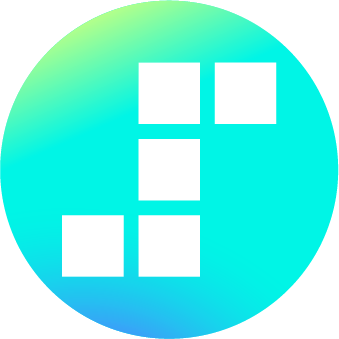}} %
StepAudio 3 Gen Technical Report
}

\author{\vspace{1em} StepFun-Audio Team
}

\renewcommand{\headeright}{StepFun-Audio Team}
\renewcommand{\undertitle}{}
\renewcommand{\shorttitle}{}

\begin{document}
\large

\raggedbottom
\setlength{\textfloatsep}{14pt plus 2pt minus 2pt}

\maketitle

\begin{abstract}

We introduce \textbf{StepAudio 3 Gen}, a general-purpose audio generation model that supports zero-shot text-to-speech (TTS), voice design, vocal generation, sound effects, music, vibe speech, and mixtures of multiple audio types within a unified framework. At its core, StepAudio 3 Gen is a \textbf{discrete autoregressive} generator that models audio directly over residual vector quantization (\textbf{RVQ}) tokens, departing from the diffusion Transformer-based continuous generation paradigm prevalent in recent general audio models. Its StepAudio Tokenizer represents general audio at \textbf{12.5~Hz} in a shared $16 \times 2048$ residual code space, jointly quantizing semantic and waveform-level acoustic features so that each code layer preserves both types of information. For generation, the backbone predicts the first codebook along the time axis using autoregressive modeling, while a lightweight causal Transformer completes the remaining fifteen codebooks along the codebook axis. Our study further identifies \textbf{three key design principles}: (1) \textbf{interference-aware progressive pretraining} for acquiring audio capabilities while preserving the textual abilities of the large language model, (2) \textbf{RVQ Adaptor} for effectively incorporating multi-codebook acoustic representations, and (3) \textbf{discrete autoregressive modeling} over a shared representation across general audio domains. With progressive pretraining, multi-task instruction training, and supervised fine-tuning, StepAudio 3 Gen achieves \textbf{state-of-the-art} performance on both \textbf{TTS} and \textbf{voice design}, while retaining strong generation capabilities across speech, vocals, sound effects, and music. Audio samples are available at \url{https://stepaudiollm.github.io/step-audio-3-gen/}.

\end{abstract}
\section{Introduction}
\label{sec:Introduction}

Audio generation is increasingly expected to cover the full range of sounds that appear in
real applications: intelligible and expressive speech, environmental sounds and sound effects,
music, and singing.  These domains have traditionally progressed along separate tracks.  Text-to-speech (TTS)
systems optimize linguistic fidelity, speaker similarity, and prosodic control
\cite{wang2023valle,guo2022prompttts,yang2023instructtts}; text-to-audio systems synthesize
non-speech events and acoustic scenes from captions \cite{kreuk2022audiogen,liu2023audioldm}; and
text-to-music and singing systems focus on musical structure, timbre, and long-range coherence
\cite{agostinelli2023musiclm,copet2023musicgen,liu2021diffsinger}.  Specialization has produced
strong models in each domain, but it also leaves applications that need several kinds of audio
with incompatible representations, conditioning formats, and generation pipelines.

Recent work has therefore moved toward unified audio generation.  One family models audio in a
continuous latent space and uses diffusion or flow matching to synthesize speech, sound, music,
or their mixtures \cite{vyas2023audiobox,tian2025audiox,xu2025uniflowaudio,dai2026qwenaudio3gen}.
Another family converts audio into discrete units and applies language-model-style sequence
modeling across tasks and modalities
\cite{borsos2023audiolm,yang2023uniaudio,zhan2024anygpt,yang2026uniaudio2}.  Continuous models
offer an effective route to parallel acoustic rendering, whereas discrete models make audio
compatible with the vocabulary, causal objective, and interleaved context of a language model.
The latter property is especially attractive when the goal is not only to generate several
audio domains, but also to place audio understanding, generation, and text intelligence in one
model.  Most closely related, UniAudio~2.0 combines a text-based large language model (LLM), factorized reasoning and
reconstruction tokens, specialized layers, a local autoregressive decoder, and multi-stage
training \cite{yang2026uniaudio2}.  It separates reasoning from reconstruction and uses a
flow-based reconstruction decoder; we study one semantic-acoustic hierarchy based on residual vector quantization (RVQ) with a shared
backbone and discrete acoustic prediction.

High-fidelity RVQ represents each frame with multiple codebook IDs.  Flattening them along time
makes the sequence prohibitively long, while using only a coarse semantic token loses acoustic
detail.  Two organizations avoid naive flattening.  Time-depth models use a temporal Transformer
for cross-frame structure and a smaller local Transformer for the codebooks within each frame
\cite{lee2022rqtransformer,yang2023uniaudio,defossez2024moshi}.  Delayed patterns instead shift
parallel codebook streams by different temporal offsets, allowing one Transformer to predict all
RVQ layers without lengthening the sequence, as in MusicGen \cite{copet2023musicgen}.  In our
setting, however, this would require a multi-stream audio output interface and make the pretrained
LLM directly predict every residual layer and receive all acoustic losses.  We choose time-depth
modeling so the LLM handles the semantically concentrated first codebook and long-range planning,
while a small module contains residual acoustic modeling and its gradients.

Even with this decomposition, adding audio tokens to a text LLM is not neutral.  The sum of many
newly initialized RVQ embeddings need not match the statistics of pretrained text embeddings,
and the residual-codebook objective supplies many acoustic predictions for every temporal
decision.  Joint optimization can force the backbone to absorb both effects before the audio
modules become useful, trading inherited language intelligence for audio capability.  Prior
audio LLMs address related forgetting risks through staged alignment, frozen modules, or
text-data rehearsal \cite{zhang2023speechgpt,wang2024freezeomni,defossez2024moshi,yang2026uniaudio2};
our design targets these two interference channels explicitly.

In this report, we present a unified audio language model built on StepAudio Tokenizer, a
12.5-Hz residual vector-quantized tokenizer shared by speech, environmental sound, music, and
singing.  Each frame is represented by 16 codebooks of size 2,048.  The tokenizer jointly
quantizes semantic and acoustic features, while semantic distillation and quantizer dropout
encourage information useful for temporal planning to concentrate in the early codebooks.  The
first codebook is incorporated into the LLM vocabulary and predicted autoregressively along the
time axis.  At every generated audio frame, the corresponding LLM hidden state and the first
codebook ID condition a four-layer Transformer, which autoregressively predicts the 15 residual
codebooks along the depth axis.  The complete RVQ frame is then decoded by the neural codec.
Thus, text and audio share a causal sequence model, and acoustic detail is generated entirely in
the discrete code space without a diffusion or flow-matching acoustic renderer.

We address interference at both the representation and optimization levels.  On the input side,
the summed embeddings of the complete RVQ frame pass through a token-wise, zero-initialized
residual module, termed the RVQ Adaptor, before being added to the ordinary token embedding.  The RVQ Adaptor gives audio
inputs a modality-specific transformation into the LLM input space without inserting a separate
sequence encoder or replacing the shared backbone.  On the optimization side, we use a
four-stage pretraining curriculum.  We first align the audio input under a frozen backbone,
then learn audio understanding jointly with replayed text.  Generation is introduced while the
conditioning hidden states of the randomly initialized residual-code predictor are detached
from the LLM, preventing its 15-codebook loss from immediately reshaping the backbone.  Only
after the predictor has converged do we restore end-to-end gradients during a low-learning-rate,
long-context cool-down.  From the second stage onward, text occupies half of each optimizer step.
A subsequent multi-task instruction-tuning stage broadens generation from speech and spoken
interaction to caption-conditioned sound, music, and singing.

The system is controlled through a unified instruction format that organizes each request into three fields: \texttt{ROLE} defines the speaker identity and voice characteristics; \texttt{DIRECTOR} describes the acoustic scene and generation intent; and \texttt{SCRIPT} arranges speech and sound events along the time axis, prefixing each spoken segment with a speaker tag and optionally a \texttt{(description)} marker, while sound effects and music are denoted as \texttt{[description]} entries to make the relative ordering of events explicit. This design enables a single instruction to define multiple characters and their relationships, specify timbre, speaking style, emotion, accent, and paralinguistic features such as laughter, breathing, and pauses, while also supporting the generation and temporal arrangement of sound effects, environmental sound, and background music.

Our main contributions are as follows:
\begin{itemize}
    \item \textbf{Interference-aware progressive pretraining.} We develop a four-stage recipe
    that progressively introduces audio alignment, understanding, generation, and end-to-end
    acoustic conditioning.  Frozen-backbone alignment, 50\% text replay, and temporary gradient
    detachment at the residual code predictor are combined to acquire audio capabilities while
    retaining the inherited textual capabilities of the LLM.

    \item \textbf{RVQ Adaptor.} We introduce a zero-initialized, token-wise
    residual adaptor that reconciles multi-codebook audio embeddings with the pretrained
    token-embedding space.  It enables the model to consume the full acoustic representation
    needed for understanding and continuation while limiting the burden on the shared backbone.

    \item \textbf{Discrete autoregressive modeling across general audio.} We build a shared LLM based generator for speech, singing, music, sound effects, and mixed audio using a single 12.5-Hz, 16-codebook RVQ representation. A time–depth architecture separates temporal prediction from within-frame codebook prediction, generating audio entirely in discrete code space without a diffusion- or flow-based acoustic renderer. 
\end{itemize}
\section{Model Architecture}
\label{sec:architecture}

Our model has two components: the StepAudio Tokenizer, which discretizes audio from the supported domains into 12.5 Hz RVQ
codes and reconstructs 24 kHz waveforms from them, and an LLM backbone that treats
these codes as an extension of its vocabulary and models text and audio in one autoregressive stream.
Generation is thus next-token prediction over the joint text--audio sequence, with the tokenizer's
decoder turning the predicted codes back into sound.

\begin{figure}[t]
\centering
\includegraphics[width=\linewidth]{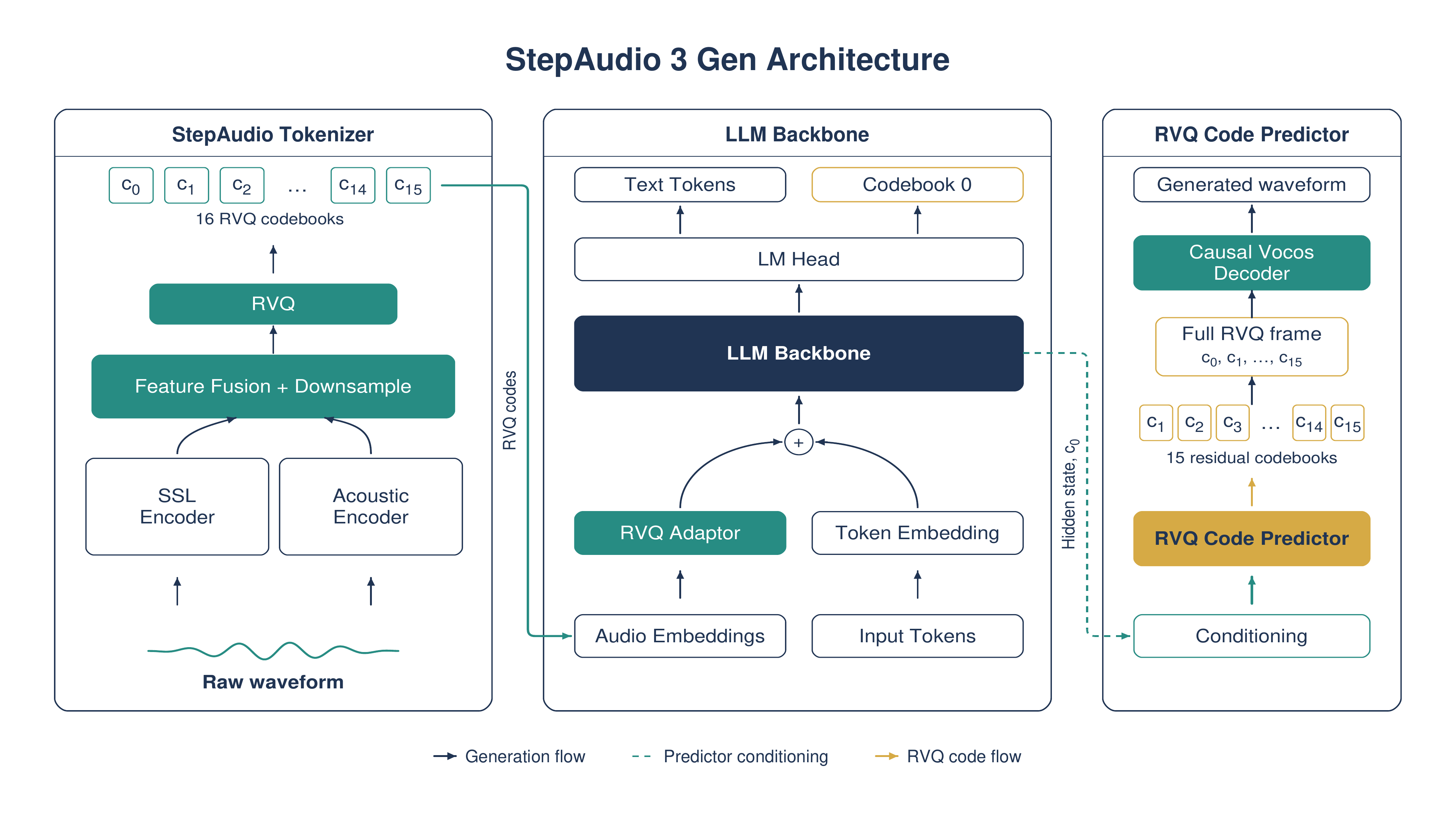}
\caption{Overview of the StepAudio 3 Gen architecture. At each audio frame, the LLM predicts $c_0$, and the RVQ Code Predictor conditions on the LLM hidden state and $c_0$ to predict $c_1,\ldots,c_{15}$. All sixteen codes form the full RVQ frame used for waveform decoding. $\oplus$ denotes element-wise addition: at audio positions, the RVQ Adaptor output is added to the corresponding token embedding (\S\ref{subsec:lm}).
}
\label{fig:model-arch}
\end{figure}

\subsection{StepAudio Tokenizer}
StepAudio Tokenizer is a 12.5 Hz audio tokenizer that discretizes speech, music,
and general audio into a single shared multi-codebook space and reconstructs waveforms at 24 kHz. Following the X-Codec
line of work \cite{ye2025xcodec,ye2025llasa}, we guide the codec toward a balance between semantic and
acoustic information by quantizing the two jointly: a
frozen self-supervised learning (SSL) encoder provides semantic features,
while a convolutional encoder with SnakeBeta activations \cite{lee2023bigvgan} extracts acoustic
features directly from the raw waveform at 50 Hz; the two representations are fused along the channel
axis, temporally compressed to 12.5 Hz by a strided convolution, and discretized by a single shared
quantizer, so that every code layer carries semantic content and acoustic detail at once rather than
being dedicated to a single modality.
Quantization is performed by a $16 \times 2048$ residual vector quantizer --- 16 layers with 2048
entries per codebook, using factorized, cosine-similarity codebook lookup \cite{kumar2023dac}. 

To enable streaming
synthesis, the decoder is fully causal: a Vocos-style Transformer backbone \cite{siuzdak2024vocos} with rotary position embeddings (RoPE) and 25-frame sliding-window attention feeds an inverse short-time Fourier transform (ISTFT) head that
emits 24 kHz audio, so the waveform is reconstructed incrementally from tokens under a bounded
receptive field and without look-ahead, supporting low-latency real-time online services. Training objectives and the two-stage recipe are described in \S\ref{subsec:tokenizer-training}.

\subsection{LLM Backbone}
\label{subsec:lm}
We retain the pretrained decoder-only Transformer architecture while extending its token embeddings and output vocabulary to support audio. Text and audio share a single autoregressive stream: the coarsest RVQ codebook (the
0th codebook) is promoted into the main language-model vocabulary as 2048 contiguous audio tokens, so
a single language-model (LM) head predicts natural-language tokens and the top-level audio code alike and the two
modalities can be interleaved freely.

On the input side, an audio frame is not read from a single vocabulary embedding. Each of its 16
codebooks (cb0 through cb15) has its own embedding table, and the 16 looked-up vectors are summed
into one frame embedding, which then passes through the RVQ Adaptor --- a token-wise stack
of residual blocks with pre-normalization and Swish-gated linear units (SwiGLU) that maps the summed audio embeddings into the LLM input space. Its output is added element-wise, at audio
positions only, to the token embedding of the codebook-0 audio token; codebook~0 is therefore
represented twice on the input side --- once in the expanded main vocabulary (as the audio token)
and once through its own table inside the summed frame.
Text positions carry no audio codes, so they receive nothing from this pathway and are left
untouched. This position gating confines the added audio pathway to audio positions.

Of the 16 codebooks in an audio frame, the LM head emits only the first; the remaining 15 residual codebooks are
produced by the RVQ Code Predictor --- a lightweight causal Transformer operating along the codebook axis. It is given two prefixes: the backbone hidden state at the
audio position (linearly projected to the predictor width) and codebook~0, the code the main LM has
just chosen; from these it autoregressively emits codebooks 1 through 15, feeding each decoded code
back as input. A stop-gradient toggle on the hidden-state prefix lets the residual-acoustic objective
train the predictor alone during the main generation stage and then propagate into the backbone
during the final joint cool-down. For waveform reconstruction, codebook~0 is retained and combined with the fifteen predicted residual codes to form the complete 16-codebook RVQ frame. In Figure~\ref{fig:model-arch}, $c_k$ abbreviates $c_{t,k}$ for a single frame $t$.

Writing $c_{t,0},\ldots,c_{t,15}$ for the sixteen codebooks of frame $t$ and $\mathbf{z}_t$ for the
backbone hidden state at that position, the generation path factorizes as
\begin{equation}
\label{eq:factorization}
\scalebox{0.9}{$\displaystyle
p(c_{t,0},\ldots,c_{t,15} \mid \mathbf{z}_{<t}, \mathbf{c}_{<t})
\;=\;
\underbrace{p(c_{t,0} \mid \mathbf{z}_{<t}, \mathbf{c}_{<t})}_{\text{main LM head}}
\;\prod_{k=1}^{15}\;
\underbrace{p(c_{t,k} \mid \mathbf{z}_t,\, c_{t,0},\ldots,c_{t,k-1})}_{\text{RVQ code predictor}}
$}
\end{equation}
The factorization is exact rather than an approximation of the joint over one frame, because the
earlier codebooks of past frames do not reach the predictor through a separate path: all sixteen
looked-up vectors are summed into the frame embedding of \S\ref{subsec:lm}, so $\mathbf{z}_t$
already encodes $\mathbf{c}_{<t}$ through the backbone. What the factorization does give up is the
reverse direction --- the predictor's output does not feed back into the backbone --- which is why
the two halves are trained jointly only once the predictor has converged.

Training optimizes the sum of a token-level term over text and codebook~0 and a term over the
fifteen residual codebooks:


\begin{equation}
\label{eq:objective}
\scalebox{0.9}{$\displaystyle
\mathcal{L}
\;=\;
\mathcal{L}_{\mathrm{tok}}
\;+\;
\lambda \, \mathcal{L}_{\mathrm{sp}},
\qquad
\mathcal{L}_{\mathrm{tok}} = -\sum \log p(c_{t,0}, x_t),
\qquad
\mathcal{L}_{\mathrm{sp}} = -\sum_{t}\sum_{k=1}^{15} \log p(c_{t,k} \mid \cdot),
$}
\end{equation}

where $\mathcal{L}_{\mathrm{sp}}$ sums rather than averages over depth, and $\lambda$ is the only
weight changed between stages (\S\ref{sec:pretrain}). Because codebook~0 occupies the model's own
vocabulary, its loss and accuracy are read off the main LM head and not off the predictor:
\begin{equation}
\label{eq:invariant}
\scalebox{0.9}{$\displaystyle
c_{t,0} \in \{0,\ldots,2047\}
\quad\Longleftrightarrow\quad
\mathrm{id}(c_{t,0}) \in [V,\, V + 2048),
$}
\end{equation}
with $V$ the text vocabulary size, which is what lets the two modalities share one softmax.
\section{Data}
\label{sec:data}

Our training data is organized into two stages: broad-coverage pretraining and task-oriented post-training. Pretraining combines text with diverse speech, vocal, music, and sound data to establish unified audio understanding and generation capabilities, while post-training uses carefully curated, task-specific corpora to further improve generation quality, naturalness, and controllability across different audio domains.

\subsection{Pretraining Data}
To introduce general-audio generation without eroding the backbone's textual competence, we build a generation-centric mixture while retaining the text corpus. Its audio sources span speech, singing, music, and environmental sounds, organized into TTS, singing synthesis, lyrics-conditioned music, caption-conditioned sound generation, and speech-to-speech translation. We also include automatic speech recognition (ASR), speech-to-text translation, audio captioning, paralinguistic question answering (QA), accent/dialect identification, and interleaved multi-turn speech dialogue. These understanding tasks anchor RVQ representations in language and expose controllable attributes such as timbre, emotion, and accent. Audio is encoded at 12.5~Hz with 16 RVQ codebooks of 2{,}048 entries each; generation samples interleave one text token with two audio frames. Across four stages, training consumes approximately 2.7T LLM tokens. From Stage~2 onward, text supplies half of each optimizer step, with the generation stages using a $3\!:\!1\!:\!2$ mixture of text, TTS, and interleaved dialogue.

\subsection{Post-training Data}

\subsubsection{Speech to Audio}

\paragraph{Text to Speech.}

To bring TTS closer to the way people naturally speak, we place additional emphasis on natural conversational speech in our supervised fine-tuning (SFT) data. While StepAudio 2.5 TTS ~\cite{lin2026stepaudio25} emphasizes global and inline instruction control, StepAudio 3 Gen focuses more on the tone, rhythm, and spontaneous vocal behaviors found in real conversations. To this end, we employ our StepAudio R1.5~\cite{zhang2026stepaudior15} for TTS-oriented audio captioning, describing the recording context, noise level, and speaker characteristics to guide the selection of natural conversational speech. We also introduce an ASR model optimized for recognizing paralinguistic phenomena to transcribe and verify candidate audio while preserving expressive cues such as tone and hesitation. Through this pipeline, we curate 1,500 hours of speech data for SFT to elicit more human-like delivery, helping the model learn natural speaking patterns conditioned on the surrounding text.


\paragraph{Speech to Vocal.}

To address the scarcity of large-scale paired speech–singing data, we construct pseudo-paired training data for speech2vocal using two complementary approaches, both designed to match speaker timbre across speech and singing. In the first approach, TTS-based voice cloning, we use the StepAudio TTS model to generate speech conditioned on a singing reference, preserving the reference timbre. In the second, singing voice conversion (SVC), we convert singing to the timbre of a target speaker and pair it with that speaker’s speech, expanding the coverage of speakers and timbres. We then filter the resulting pairs using the Production Quality (PQ) and Content Enjoyment (CE) scores from Audiobox Aesthetics~\cite{tjandra2025audioboxaesthetics}, singing mean opinion score (MOS) predictions from SingMOS~\cite{singmosofficial}, and the timbre similarity between synthesized and target audio, retaining only samples that meet both audio-quality and timbre-similarity criteria.
  

\subsubsection{Text to Audio}
\label{subsec:unified-training-format}

\paragraph{Voice Design.}
We construct the voice-design SFT corpus from recorded and synthesized speech, covering a wide range of expressive styles and acoustic conditions. The recorded portion includes both natural conversational and performed speech, spanning everyday and spontaneous conversations, character dialogue, and professional voice-over across diverse recording environments. We further incorporate synthesized speech to expand coverage of underrepresented combinations of voice characteristics, speaking styles, and acoustic scenarios, including conversation, narration, and speech–music mixtures.


\paragraph{Vocal data.}
We first separate full songs, remove tracks that are speech-dominated, contain little singing, have low lyric-alignment coverage, or fail overall quality checks, and extract candidate segments at lyric-line boundaries using multiple gating criteria. We then evaluate the candidate segments produced by the two separation routes using Audiobox Aesthetics PQ/CE scores, SingMOS predictions, loudness, and spectral bandwidth. We cross-check their quality scores and ASR transcriptions and apply joint thresholds to obtain the final vocal dataset.

\paragraph{Music data.}
For music generation, we construct a text-conditioned instrumental music corpus covering diverse scenarios, moods, genres, instrumentations, rhythmic patterns, and production styles. We use the StepAudio Music Model to generate instrumental tracks from text prompts specifying the target musical content and intended use cases. For each generated track, we then create diverse Chinese and English textual conditions, including concise descriptions, detailed descriptions, and conversational requests. Detailed descriptions emphasize instrumentation, rhythm, and production style, whereas conversational requests focus on user preferences and usage contexts. This multi-form prompt design provides training supervision across different levels of musical description and user intent.


\paragraph{Sound data.}
To construct large-scale and diverse training data for sound-effect generation, we combine sound-effect recordings obtained from diverse corpora and sound-effect libraries with synthesized speech-sound mixtures. We standardize heterogeneous metadata and enrich the recordings with multi-level annotations covering sound events, sources, acoustic attributes, scenes, and temporal structure. We retain samples with usable source-level descriptions and filter out recordings containing speech, personally identifiable information, or other unsuitable content. For the retained recordings, we preserve source and licensing metadata where available. For speech-sound mixtures, we sample scene skeletons specifying speakers, speech segments, sound effects, and ambience, and use our text-based large language model to complete the speech content, speaking styles, sound descriptions, durations, and temporal relations. Speech and non-speech sounds are then generated as separate stems using dedicated synthesis models. Finally, we remove failed or potentially inaudible sources, normalize and align the remaining stems, and mix them with gains adjusted by source prominence and distance to preserve the intended foreground–background structure.

\section{Training}
\label{sec:training}

Training proceeds in two parts. We first train the StepAudio Tokenizer to obtain a compact and semantically grounded discrete representation that preserves both linguistic structure and acoustic detail. We then initialize from a pretrained text LLM and progressively introduce audio understanding and generation through a four-stage pretraining recipe designed to integrate the audio modality while minimizing interference with the backbone's inherited textual capabilities.

\subsection{Tokenizer Training}
\label{subsec:tokenizer-training}

We train the RVQ with a quantizer dropout \cite{zeghidour2022soundstream}
of 0.5, which drives the lower layers to encode coarse-grained, semantically aligned information while
the deeper layers progressively refine the details left behind; the ratio is deliberately moderate, as
a larger dropout further strengthens the coarse layers but slightly degrades full-depth reconstruction.
Semantic grounding is enforced by a distillation branch that regresses the SSL encoder teacher features
from the quantized latent; since this objective is applied under quantizer dropout, semantically
aligned information is concentrated in the earliest codebooks.

Training adopts a generative adversarial network (GAN) framework \cite{zeghidour2022soundstream,defossez2023encodec} in which the
generator operates end-to-end on raw waveforms to extract, quantize, and resynthesize audio, while a
multi-period waveform discriminator \cite{kong2020hifigan} and a multi-resolution spectral
discriminator improve the naturalness and fidelity of reconstructed audio, complemented by
feature-matching, multi-scale mel-spectrogram \cite{kumar2023dac}, and residual-quantization commitment
losses that enforce time--frequency consistency and stable codebook usage.

The tokenizer
is trained in two stages: large-scale pretraining on roughly 700k hours of speech, music,
and sound events with a modality-balanced sampler, followed by a decoder-only refinement stage in which
the encoder, quantizer, and semantic branch are frozen and only the acoustic decoder and discriminators
are updated -- improving reconstruction quality while leaving the token space untouched, so models
already trained on these tokens remain compatible.

\subsection{LLM Pretraining}
\label{sec:pretrain}

\begin{table}[t]
\centering
\small
\caption{Four-stage pretraining recipe. Batch size is measured in tokens per optimizer step,
where one temporal LLM step counts as one token. Thus, audio represented at 12.5~Hz contributes
12.5 tokens per second, although each audio token contains 16 RVQ codebook IDs. LR denotes learning rate.}
\label{tab:stages}
\begin{tabular}{llrll}
\toprule
Stage & Focus & Batch Size (tokens) & LR & LR scheduler \\
\midrule
1 & modality alignment        & $4.19$M  & $2\times10^{-4} \to 2\times10^{-5}$ & cosine \\
2 & audio understanding       & $12.58$M & $2\times10^{-5}$ & constant \\
3 & generation, detached      & $12.58$M & $2\times10^{-5}$ & constant \\
4 & long-context cool-down    & $25.17$M & $2\times10^{-5} \to 1.5\times10^{-5}$ & cosine \\
\bottomrule
\end{tabular}
\end{table}

\label{subsec:pretrain-rationale-text}
We initialize from a pretrained text LLM, which serves as the pretraining backbone, and adopt a four-stage paradigm consuming approximately 2.7T tokens.  Preserving the textual ability of the backbone is a second objective throughout: Stage~1 trains only the adaptor under a frozen backbone, Stage~3 detaches the RVQ code predictor, and from Stage~2 onward the original text corpus makes up half of every optimizer step.

\textbf{Stage~1: Modality Alignment.} The first stage aligns audio representations with a frozen LLM using ASR and speech-to-text translation data, in which only the textual output is supervised. We train only the input-side audio embedding and adaptor, holding the backbone, the LM head, and the code predictor at zero learning rate, while the token embedding receives a $0.1\times$ multiplier so that the newly added audio-code rows can adapt. As the down projections of the adaptor are zero-initialized, the module is an exact identity map at initialization; together with the frozen backbone this makes the drift in textual ability strictly zero, so no text replay is needed here. 

\textbf{Stage~2: Audio Understanding.}
\label{subsec:pretrain-s2}
The full model is then unfrozen and trained on audio-understanding tasks mixed one-to-one with the text corpus, still without supervising any audio target. The learning rate drops to a constant $2\times10^{-5}$, forming a plateau that later stages continue from without re-warming, while the from-scratch modules retain a $10\times$ multiplier and are excluded from weight decay. 

\textbf{Stage~3:  Generation with a Detached Code Predictor.} Stage~3 is the main pretraining run and introduces generation, with the mixture set to text\,:\,TTS\,:\,interleaved dialogue $=3\!:\!1\!:\!2$ so that the text share stays at $50\%$ and TTS accounts for $1/6$ of every micro-batch, or one third of the audio-generation streams. Here the conditioning input of the code predictor is detached from the backbone hidden states: supervision over the 15 residual codebooks is far stronger in magnitude than the token-level objective, yet the predictor is still randomly initialized, so its gradient would reshape the representation built over the previous two stages. Once detached, the residual acoustic part trains the predictor alone while the backbone stays driven by the token-level objective, which decouples learning acoustic detail from learning the semantic and prosodic plan and serves as a second safeguard for textual ability.  Stage~3 uses $\lambda = 1.0$. Because $\mathcal{L}_{\mathrm{sp}}$ sums rather than averages over the fifteen residual codebooks, the acoustic term can be substantially larger than the token-level term. Detachment prevents it from directly updating the backbone in this stage; after detachment is removed, the same weighting can cause it to dominate backbone updates.

\textbf{Stage~4:  Long-Context Joint Cool-down.}
\label{subsec:pretrain-s4}
The detachment is finally removed, so the residual acoustic part propagates into the backbone and its hidden states are shaped to serve as good acoustic conditioning as well. This is deferred to last because the predictor is by now well converged and because the low cool-down learning rate limits the extent to which its gradients alter the backbone representations. The same run lowers $\lambda$ from $1.0$ to $0.1$: with the depth loss now reaching the backbone, the two terms are brought to comparable magnitude rather than leaving the sum over fifteen codebooks an order of magnitude above the token-level objective. We extend the sequence length from $16{,}384$ to $32{,}768$ using context parallelism, which at 12.5~Hz covers roughly 44 minutes of audio and suffices for long-form synthesis and multi-turn dialogue. 


\subsection{Post-training}
\textbf{Supervised Fine-tuning.}
Building on the pretrained model, we run full-parameter fine-tuning to further improve its instruction-following and speech-generation abilities. The joint configuration from the end of pretraining is carried over: the detachment stays off and the speech loss, kept at $\lambda = 0.1$, continues to propagate through the code predictor into the backbone. The SFT corpus contains approximately 5{,}000 hours of audio, spanning music, sound effects, voice design, singing, and speech. The music and text-conditioned singing subsets are generated with StepAudio Music Model. The audio embedding, the adaptor, and the code predictor retain a $10\times$ learning-rate multiplier, while the remaining parameters are fine-tuned at the base rate. Training uses a sequence length of 16{,}384 and a batch size of 64, with a cosine schedule from $1.0\times10^{-5}$ to $1.0\times10^{-6}$.


\textbf{Reinforcement Learning.} After supervised fine-tuning, we apply Group Relative Policy Optimization
(GRPO)\cite{shao2024deepseekmathpushinglimitsmathematical} to further improve instruction following. The training data follow the
same task taxonomy and conditioning formats as the SFT data, covering TTS,
Voice Design, vocal, music, and sound generation, as well as speech-and-sound
mixtures. Speech-containing examples include an exact transcript in addition
to role and style requirements, while the other tasks use concrete descriptions
of acoustic events and performance characteristics. Ambiguous or weakly
verifiable instructions are removed to reduce reward noise.

For each generated audio $y_i$, an audio understanding model~\cite{zhang2026stepaudior15}  first produces a
caption $c_i$. A text LLM compares $c_i$ with the original instruction $x$ and
returns an instruction-consistency score between 0 and 100. We independently
score each instruction--caption pair four times and use the mean $\bar{s}_i$
to reduce evaluation variance. For speech examples with target transcript $t$,
an ASR model produces $\hat{t}_i$, from which we compute either character error rate (CER) or word error rate (WER),
denoted by $e_i$. The reward is

\begin{equation}
\scalebox{0.9}{$\displaystyle
    R_i = \frac{\bar{s}_i}{100}
    \begin{cases}
        \exp(-\tau e_i), & e_i \leq 0.5,\\
        0, & e_i > 0.5,
    \end{cases}
    \qquad
    \bar{s}_i=\frac{1}{4}\sum_{k=1}^{4}s_i^{(k)}.
$}
\end{equation}

Here, $\tau=3$ controls the exponential penalty on the recognition error. For
tasks without a target transcript, the ASR term is set to one. This
multiplicative reward prevents an output with severely incorrect linguistic
content from receiving a high score solely because its acoustic style matches
the instruction.

For every instruction, we sample $G=16$ candidate audio sequences. GRPO uses
their group-relative rewards without training a separate value model. Given
the group mean $\mu_R$ and standard deviation $\sigma_R$, the advantage is
$\hat{A}_i=(R_i-\mu_R)/(\sigma_R+\eta)$, where $\eta$ is a small constant for
numerical stability. Inspired by the dynamic sampling
strategy of DAPO~\cite{yu2025dapoopensourcellmreinforcement}, groups with $\sigma_R<\delta$ are
discarded, where $\delta=0.02$ is the minimum within-group reward standard
deviation retained for training. Such groups contain little reliable
preference signal. The policy is optimized with



\begin{equation}
\scalebox{0.85}{$\displaystyle
J_{\mathrm{GRPO}}(\theta)=\mathbb{E}\!\left[\frac{1}{G}\sum_{i=1}^{G}\frac{1}{16|y_i|}\sum_t\sum_{\ell=0}^{15}\min\!\left(\rho_{i,t,\ell}\hat{A}_i,\operatorname{clip}\!\left(\rho_{i,t,\ell},1-\epsilon_{\mathrm{clip}},1+\epsilon_{\mathrm{clip}}\right)\hat{A}_i\right)\right]-\beta D_{\mathrm{KL}}\!\left(\pi_\theta\,\|\,\pi_{\mathrm{ref}}\right),
$}
\end{equation}

where \(t\) indexes the generated audio positions and \(\ell\in\{0,\ldots,15\}\) indexes the RVQ codebook (one base codebook plus 15 residual codebooks). Here \(\rho_{i,t,\ell}\) is the token-level importance ratio between the updated policy \(\pi_\theta\) and the rollout policy for each codebook token, \(\epsilon_{\mathrm{clip}}\) is the clipping range, and \(\pi_{\mathrm{ref}}\) is the SFT reference policy.


We use a small Kullback–Leibler (KL) divergence coefficient \(\beta\), relying primarily on policy clipping to constrain updates, and train with a learning rate from \(10^{-6}\) to \(10^{-7}\).

\label{subsec:rl}

\section{Evaluation}
\label{sec:evaluation}

Our evaluation is organized around three questions: whether the proposed RVQ Adaptor effectively integrates multi-codebook audio representations into the pretrained LLM, 
whether the interference-aware progressive pretraining strategy preserves the LLM's textual capabilities,
and whether the resulting model delivers strong generation quality and controllability. We first evaluate the effectiveness of the RVQ Adaptor and the interference-aware progressive pretraining strategy, and then focus on two core generation capabilities. For TTS, we assess perceived human-likeness through subjective human evaluation; for voice design, we evaluate both instruction-following performance on an objective benchmark and human preference through subjective evaluation.

\subsection{RVQ Adaptor}
\label{subsec:exp_adapter}

We compare systems with and without the RVQ Adaptor after pretraining to evaluate the effect of aligning audio embeddings with the pretrained LLM’s token embedding space on its audio-language capabilities. The evaluation covers 1) automatic speech recognition (ASR) on the AISHELL-1 test set~\cite{aishell} and the LibriSpeech test-clean set~\cite{librispeech}, measured by character error rate (CER) and word error rate (WER), respectively; 2) audio understanding and reasoning on MMAU~\cite{mmau}, measured by accuracy; 3) bidirectional English–Chinese speech-to-text translation on CoVoST~\cite{covost}, measured by BLEU; and 4) knowledge question answering on SpeechMMLU~\cite{mimo}, measured by accuracy in the text-to-speech (T2S) setting. As shown in Table~\ref{tab:adapter_audio_comparison}, the system equipped with the RVQ Adaptor consistently outperforms its counterpart across all reported metrics, indicating stronger overall audio-language capabilities. 

\begin{table*}[htbp]
\centering
\caption{Comparison of systems with and without the RVQ Adaptor on audio benchmarks
after pretraining.}
\label{tab:adapter_audio_comparison}
\small
\setlength{\tabcolsep}{4pt}
\renewcommand{\arraystretch}{1.1}
\begin{tabular}{lcccccc}
\toprule
\multirow{2}{*}{System}
& AISHELL-1
& LibriSpeech
& MMAU
& \multicolumn{2}{c}{CoVoST BLEU $\uparrow$}
& SpeechMMLU (T2S) \\
\cmidrule(lr){2-2}
\cmidrule(lr){3-3}
\cmidrule(lr){4-4}
\cmidrule(lr){5-6}
\cmidrule(lr){7-7}
& CER (\%) $\downarrow$
& WER (\%) $\downarrow$
& Acc. (\%) $\uparrow$
& En$\to$Zh
& Zh$\to$En
& Acc. (\%) $\uparrow$ \\
\midrule
w/o RVQ Adaptor
& 5.25 & 6.00 & 40.70
& 12.05 & 5.99 & 12.13 \\
w/ RVQ Adaptor
& \textbf{3.00} & \textbf{3.41} & \textbf{51.70}
& \textbf{30.56} & \textbf{18.59}
& \textbf{58.76} \\
\bottomrule
\end{tabular}
\end{table*}

\subsection{Interference-aware progressive pretraining}
\label{subsec:exp_pretrain}

To evaluate the effectiveness of our interference-aware progressive pretraining strategy in preserving the pretrained LLM’s text-domain capabilities after pretraining, we compare it with a three-stage baseline on text benchmarks. The baseline does not use the RVQ Adaptor and follows three stages: ASR training, audio understanding and generation training, and a final cool-down stage. The evaluation assesses knowledge and reasoning capabilities using FinEval~\cite{fineval}, C-Eval~\cite{ceval}, MMLU~\cite{mmlu}, CMMLU~\cite{cmmlu}, MATH~\cite{math}, GSM8K~\cite{gsm8k}, and BBH~\cite{bbh}, together with code generation performance on HumanEval~\cite{humaneval}. We report accuracy for FinEval, C-Eval, MMLU, and CMMLU; exact match (EM) for MATH, GSM8K, and BBH; and Pass@1 for HumanEval. As shown in Table~\ref{tab:progressive_text_comparison}, our interference-aware progressive pretraining strategy consistently outperforms the baseline across all text benchmarks, indicating better retention of the pretrained LLM’s knowledge and reasoning capabilities.

\begin{table*}[htbp]
\centering
\caption{Comparison of the baseline and our interference-aware progressive pretraining
on text benchmarks after pretraining.}
\label{tab:progressive_text_comparison}
\small
\setlength{\tabcolsep}{3pt}
\renewcommand{\arraystretch}{1.1}
\begin{tabular}{lcccccccc}
\toprule
\multirow{2}{*}{System}
& FinEval
& C-Eval
& MMLU
& CMMLU
& MATH
& GSM8K
& BBH
& HumanEval \\
\cmidrule(lr){2-9}
& Acc. (\%) $\uparrow$
& Acc. (\%) $\uparrow$
& Acc. (\%) $\uparrow$
& Acc. (\%) $\uparrow$
& EM (\%) $\uparrow$
& EM (\%) $\uparrow$
& EM (\%) $\uparrow$
& Pass@1 (\%) $\uparrow$ \\
\midrule
Baseline
& 65.86 & 66.34 & 64.99 & 66.69
& 36.62 & 67.94 & 59.61 & 47.56 \\
Interference-aware
& \textbf{71.68} & \textbf{71.92}
& \textbf{69.20} & \textbf{71.73}
& \textbf{45.07} & \textbf{74.05}
& \textbf{67.03} & \textbf{54.27} \\
\bottomrule
\end{tabular}
\end{table*}

\subsection{TTS}
\label{subsec:tts-evaluation}

Given our emphasis on human-like delivery in TTS, we specifically evaluate how closely the generated speech resembles real human speech. We conduct an Arena-style pairwise evaluation, where raters judge the overall listening impression and select the sample that sounds more like a real person speaking naturally. This holistic judgment is intended to capture prosody, rhythm, expressiveness, and spontaneous speaking behaviors, with ties allowed when no clear preference exists.

We favor subjective comparison over objective metrics such as character error rate (CER) and speaker similarity (SS), which capture only limited aspects of TTS performance. A model with flat prosody can still achieve a low CER as long as the content is correctly conveyed, while SS primarily reflects speaker identity and timbre rather than expressive delivery. These metrics therefore do not adequately reflect the human-like expressiveness targeted by our model.

We construct a Chinese human-likeness evaluation set from text extracted from real human speech. We use one voice from each competing TTS model for comparison with StepAudio 3 Gen. The six systems are evaluated through Arena-style pairwise comparisons, with 1,500 comparisons in total. Each comparison is recorded as a win, tie, or loss. Based on these outcomes, we report both Elo ratings over the full evaluation pool and direct head-to-head results against five leading commercial TTS systems.

\begin{figure}[htbp]
    \centering
    \begin{subfigure}{0.49\linewidth}
        \centering
        \includegraphics[width=\linewidth]
            {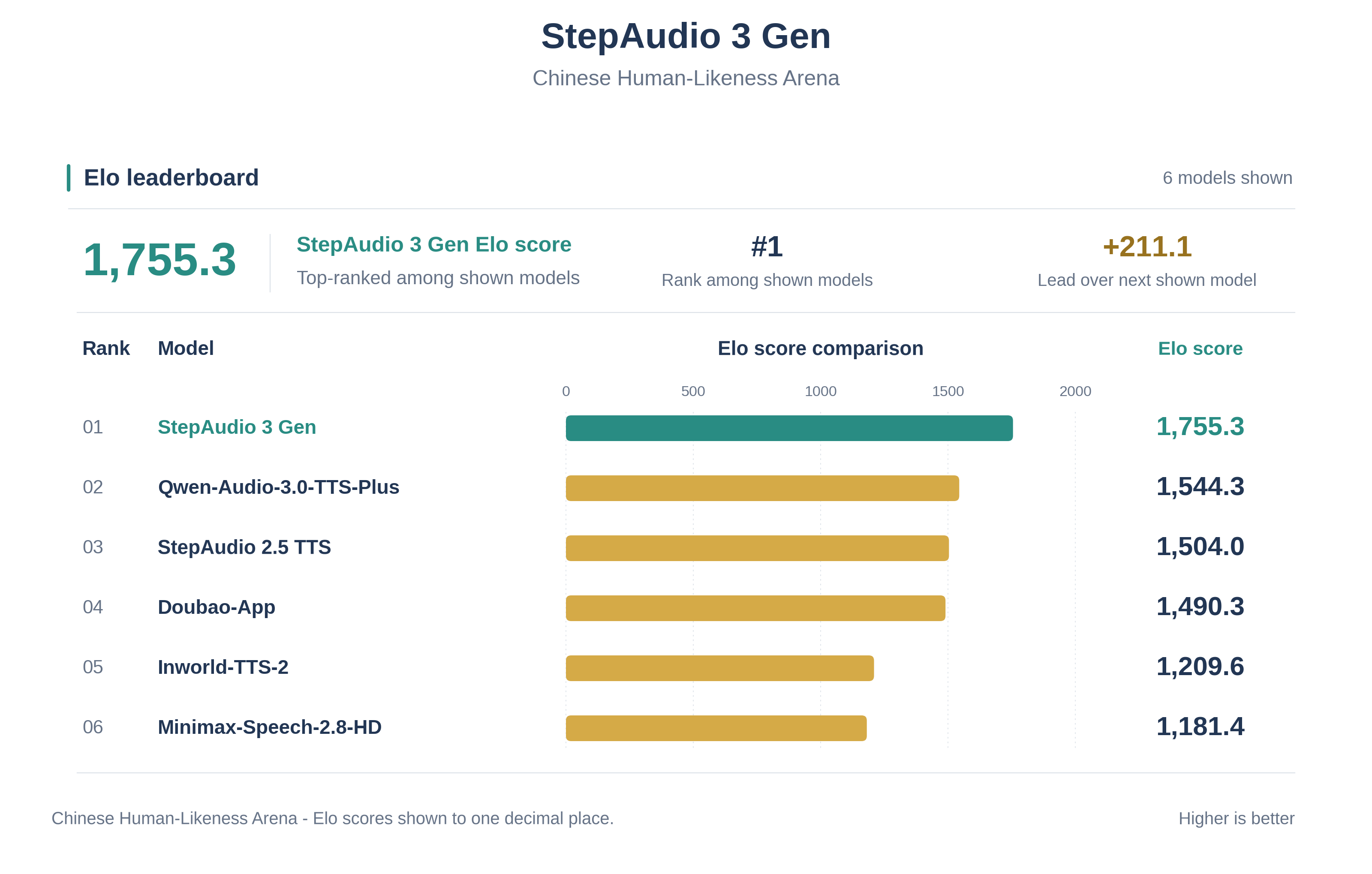}
        \caption{Elo rankings of models on the Chinese human-likeness test set under human evaluation.}
        \label{fig:tts-elo}
    \end{subfigure}
    \hfill
    \begin{subfigure}{0.49\linewidth}
        \centering
        \includegraphics[width=\linewidth]
            {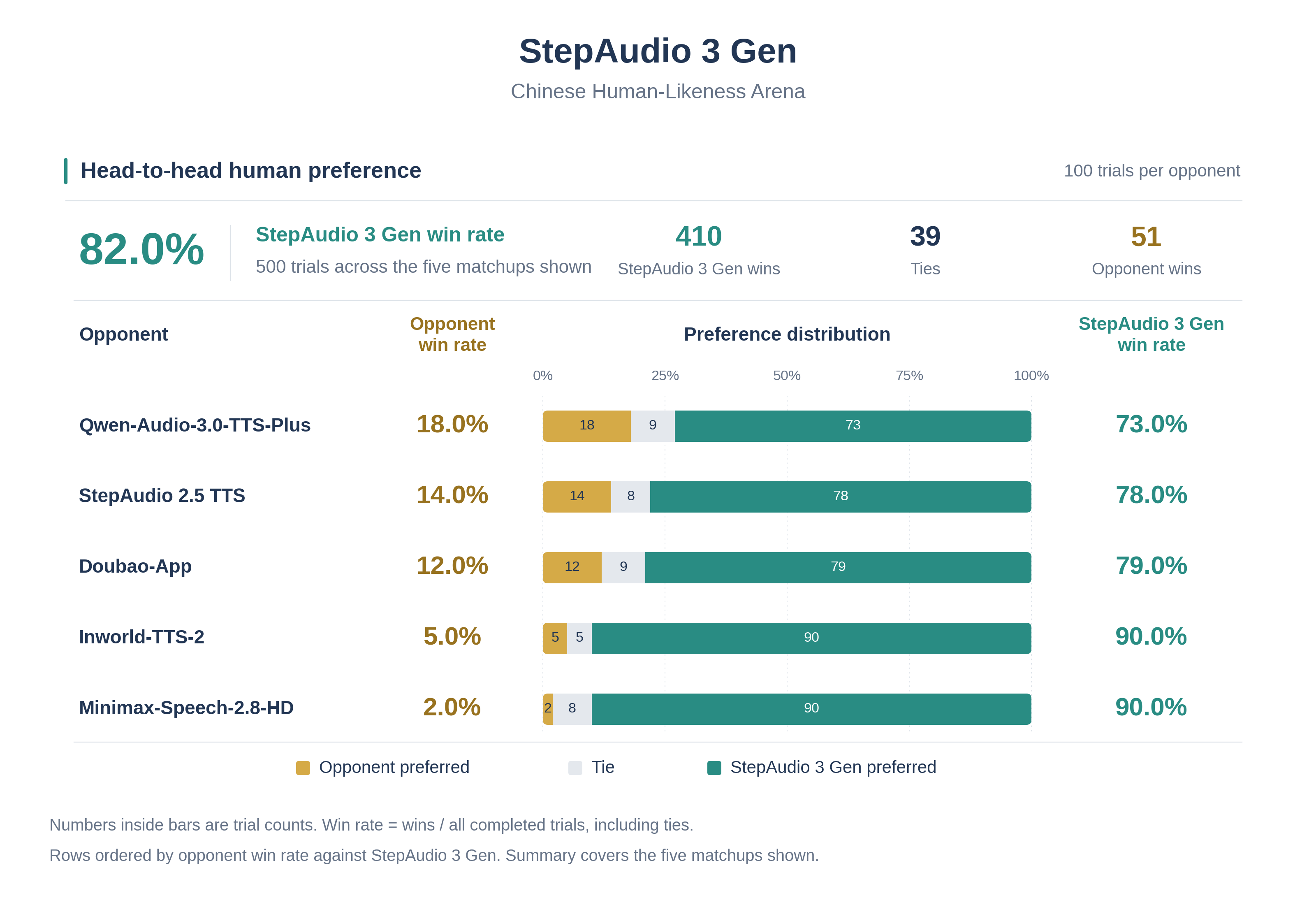}
        \caption{Head-to-head results for StepAudio 3 Gen against baseline models on the Chinese human-likeness test set.}
        \label{fig:tts-head-to-head}
    \end{subfigure}
    \caption{Human evaluation on the Chinese human-likeness test set.}
    \label{fig:tts-eval}
\end{figure}

We aggregate the pairwise outcomes using Elo ratings, with higher scores indicating stronger preference within the evaluation pool. Figure~\ref{fig:tts-eval}\subref{fig:tts-elo} shows that StepAudio 3 Gen achieves the highest Elo rating of 1755.33, outperforming five leading commercial TTS systems: Qwen-Audio-3.0-TTS-Plus~\citep{xiang2026qwenaudio3tts}, Doubao-App~\citep{bytedancedoubaoapp}, MiniMax-Speech-2.8-HD~\citep{minimaxspeech28}, Inworld-TTS-2~\citep{inworldtts2}, and StepAudio 2.5 TTS~\citep{lin2026stepaudio25}.





Figure~\ref{fig:tts-eval}\subref{fig:tts-head-to-head} further presents direct head-to-head results between StepAudio 3 Gen and five leading commercial TTS systems, with 100 comparisons per opponent. StepAudio 3 Gen achieves an aggregate win rate of 82.0\%, with win rates ranging from 73.0\% to 90.0\% across individual opponents, maintaining a clear majority in every matchup. As preference in this evaluation is defined by the overall perceived human-likeness of the generated speech, these results demonstrate a consistent advantage of StepAudio 3 Gen across all five competing systems.



\subsection{Voice Design}
\label{subsec:vd-evaluation}

\textbf{InstructTTSEval benchmark.}
We evaluate style following on the full InstructTTSEval benchmark~\citep{huang2025instructttseval}:
1{,}000 Chinese and 1{,}000 English texts, each evaluated under Acoustic-Parameter
Specification (APS), Descriptive-Style Directive (DSD), and Role-Play (RP),
for a total of 6{,}000 generated utterances per system.  Gemini~3.1~Pro~\citep{google2026gemini31pro}
(\texttt{gemini-3.1-pro-preview}) judges each audio against its original
benchmark instruction and returns a binary style-consistency decision. Scores
are the percentages of valid judgments marked consistent for each condition;
\texttt{AVG} is their unweighted mean.

The baseline systems include
Qwen3-TTS-12Hz-1.7B-VoiceDesign~\citep{hu2026qwen3tts},
MOSS-VoiceGenerator~\citep{huang2026mossvoicegenerator},
Ming-omni-tts-0.5B~\citep{inclusionaimingomnitts},
and VoiceSculptor~\citep{hu2026voicesculptor}.
Table~\ref{tab:vd-instructttseval} reports Gemini style-consistency scores on the full benchmark.
Among the evaluated systems, StepAudio 3 Gen achieves the best Chinese performance
across APS, DSD, and RP, with an average score of 85.2\%, and also attains the best English performance with an average score of 77.7\%.

\begin{table}[t]
\centering
\footnotesize
\setlength{\tabcolsep}{4pt}
\caption{StepAudio 3 Gen style-consistency scores on the full InstructTTSEval benchmark (2,000 texts under three conditions, totaling 6,000 generated utterances). Scores are percentages; higher is better. For each language, AVG is the unweighted mean of APS, DSD, and RP.}
\label{tab:vd-instructttseval}
\begin{adjustbox}{max width=\linewidth}
\begin{tabular}{lcccccccc}
\toprule
 & \multicolumn{4}{c}{\textbf{InstructTTSEval-EN}} & \multicolumn{4}{c}{\textbf{InstructTTSEval-ZH}} \\
\cmidrule(lr){2-5}\cmidrule(lr){6-9}
 & APS$\uparrow$ & DSD$\uparrow$ & RP$\uparrow$ & AVG$\uparrow$
          & APS$\uparrow$ & DSD$\uparrow$ & RP$\uparrow$ & AVG$\uparrow$ \\
\midrule
Qwen3-TTS-12Hz-1.7B-VoiceDesign
 & 74.7 & 83.8 & 61.9 & 73.5 & 79.0 & 86.7 & 56.2 & 74.0 \\
MOSS-VoiceGenerator
 & 59.5 & 79.0 & 56.1 & 64.9 & 63.7 & 78.1 & 53.4 & 65.1 \\
Ming-omni-tts-0.5B
 & 71.6 & 67.7 & 55.9 & 65.1 & 75.2 & 83.8 & 54.0 & 71.0 \\
VoiceSculptor
 & -  &  -   &  -   &  -   & 49.2 & 62.5 & 45.9    & 52.5     \\
\midrule
\textbf{StepAudio 3 Gen (Ours)}
 & \textbf{82.4} & \textbf{85.5} & \textbf{65.2} & \textbf{77.7} & \textbf{88.5} & \textbf{91.5} & \textbf{75.5} & \textbf{85.2} \\
\bottomrule
\end{tabular}
\end{adjustbox}
\end{table}



\begin{figure}[htbp]
    \centering
    \begin{subfigure}{0.49\linewidth}
        \centering
        \includegraphics[width=\linewidth]
            {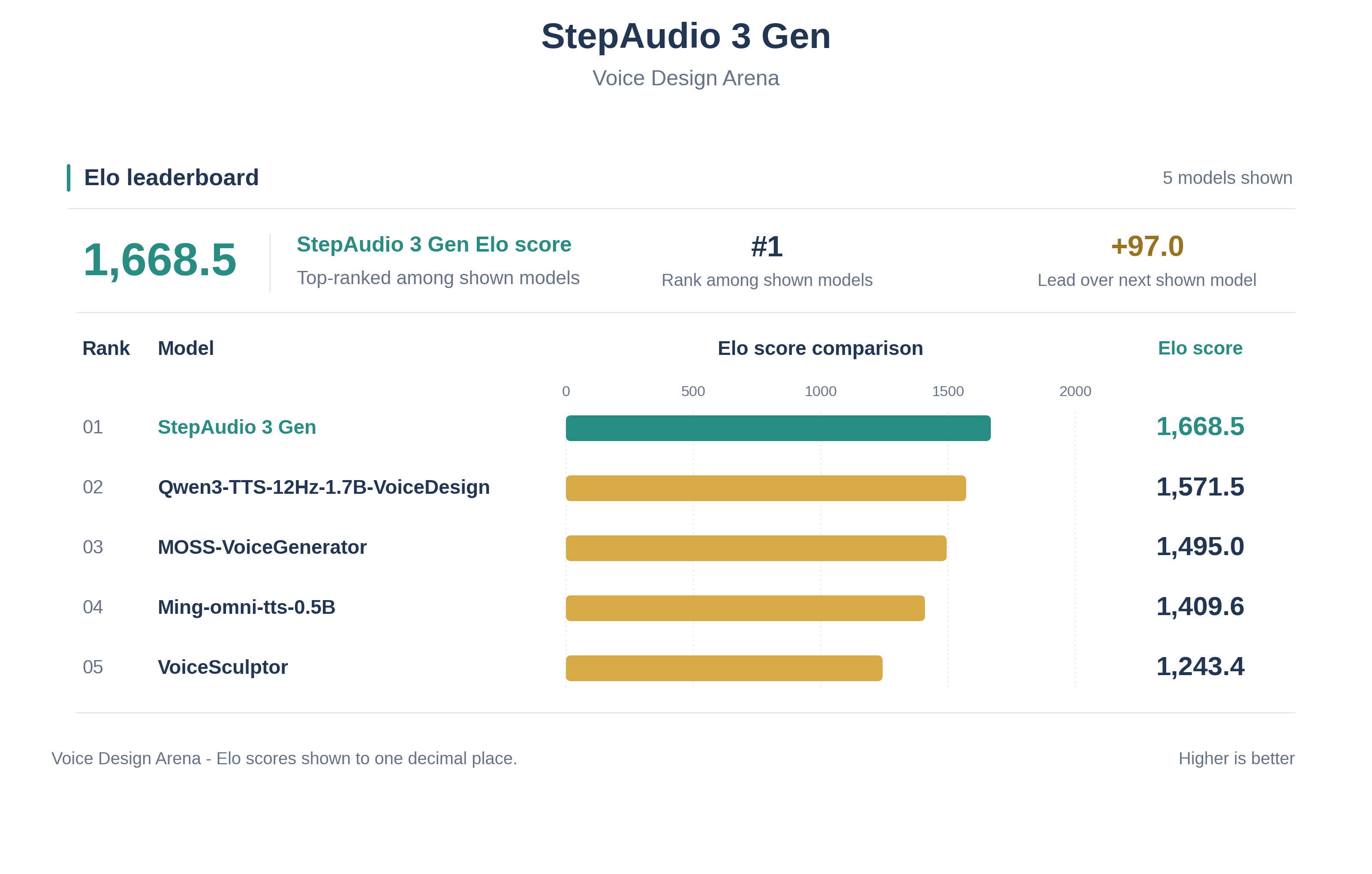}
        \caption{Elo rankings of models on the voice design task under blind human evaluation.}
        \label{fig:vd-elo}
    \end{subfigure}
    \hfill
    \begin{subfigure}{0.49\linewidth}
        \centering
        \includegraphics[width=\linewidth]
            {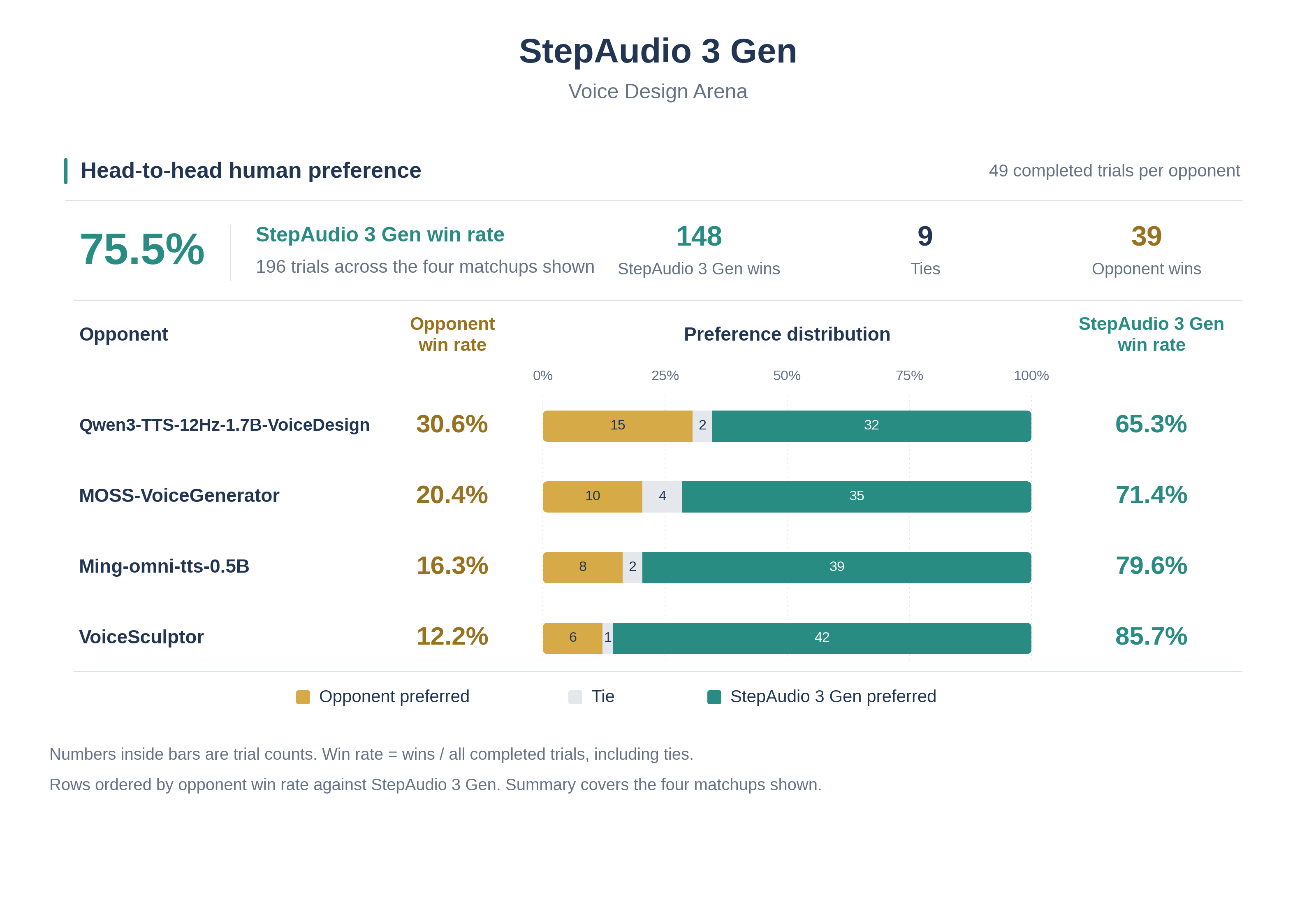}
        \caption{Head-to-head results for StepAudio 3 Gen against baseline models on the voice design task.}
        \label{fig:vd-head-to-head}
    \end{subfigure}
    \caption{Human evaluation on the voice design task.}
    \label{fig:vd-eval}
\end{figure}

\textbf{Human Preference Evaluation on Voice Design.}
We conduct a blind human preference evaluation on the voice design task to assess the models' instruction-following capability. We employ an Elo-based battle mechanism to evaluate five models with voice design capabilities. A total of 490 head-to-head comparisons covering both Chinese and English were conducted across all models, with Elo scores dynamically calculated based on match outcomes.
As shown in Figure~\ref{fig:vd-elo}, StepAudio 3 Gen ranks first with an Elo score of 1668.5, demonstrating superior human preference performance on the voice design task compared to current mainstream models. Figure~\ref{fig:vd-head-to-head} further presents direct head-to-head results for StepAudio 3 Gen against baseline models on the voice design task. StepAudio 3 Gen achieves an aggregate win rate of 75.5\%, with win rates ranging from 65.3\% to 85.7\% across individual opponents. As preference in this evaluation is defined by overall listening impression and instruction-following quality, these results demonstrate a consistent advantage of StepAudio 3 Gen on the voice design task.
\section{Extensions}
\label{sec:evaluation}

StepAudio 3 Gen models speech and general audio within a single shared RVQ token space. The same discrete autoregressive framework supports the TTS and voice design capabilities evaluated above, as well as vocals, music, sound effects, and complex audio scenes presented in this section. All of these capabilities are modeled and generated entirely in the shared discrete code space, without introducing task-specific tokenizers or diffusion- or flow-based acoustic renderers. By jointly training the RVQ representation and generator across diverse audio domains, the model learns semantic and acoustic structures beyond speech and develops solid general-audio generation capabilities. We highlight several representative capabilities below.

\paragraph{Vocal Generation.}
StepAudio 3 Gen supports both text to vocal and speech to vocal generation. For text to vocal generation, users can specify the target voice, lyrics, musical key, and vocal style through natural language prompts, allowing the model to jointly control vocal identity, linguistic content, pitch characteristics, and expressive delivery. For speech to vocal generation, the input speech serves as a reference for the target voice, while the lyrics, musical key, and singing style are specified separately to generate singing in the reference speaker's voice.

\paragraph{Music Generation.}
The model supports long-form instrumental music generation from natural language prompts describing genre, mood, instrumentation, rhythmic characteristics, production style, and intended use. StepAudio 3 Gen can generate coherent musical pieces exceeding 60 seconds while maintaining overall musical structure and stylistic consistency over extended durations.

\paragraph{Sound Generation.}
StepAudio 3 Gen supports both isolated sound-effect generation and complex acoustic-scene generation. It can synthesize individual sound events from concise descriptions, while also handling more complex prompts involving multiple sound sources, layered events, ambience, and interactions among acoustic events. The model further supports temporal arrangement of sound events, allowing prompts to specify their ordering, relative timing, and co-occurrence within a scene.

\paragraph{Vibe Speech.}
StepAudio 3 Gen can generate vibe speech within composite acoustic scenes, jointly modeling spoken content, speaking style, atmosphere, background sound, and contextual cues. It also supports multi-speaker dialogue and interactions among multiple speakers and sound sources, enabling the generation of complex conversational scenes with coherent acoustic environments.
\section{Conclusion}
\label{sec:conclusion}

We present StepAudio 3 Gen as a step toward unified general-purpose audio generation within a discrete autoregressive framework. Rather than relying on separate generation paradigms for different audio domains, our results show that a shared semantic--acoustic representation can support diverse audio generation tasks within a single LLM-based architecture. Our experiments further suggest three practical lessons: audio capabilities should be introduced progressively to mitigate interference with the pretrained LLM, multi-codebook acoustic representations can be effectively integrated through a lightweight RVQ Adaptor, and discrete autoregressive modeling can scale beyond speech to general audio without a continuous acoustic renderer. StepAudio 3 Gen achieves state-of-the-art results on TTS and voice design, while also demonstrating solid generation capabilities in vocals, music, sound, and vibe speech. Taken together, these findings demonstrate that discrete autoregressive modeling provides a practical and scalable alternative to prevailing continuous-generation approaches for building unified audio models.

\clearpage
\section*{Contributors}

\noindent
Contributors are listed alphabetically by first name.

\vspace{0.8em}

\setlength{\columnsep}{2.5em}
\begin{multicols}{4}
\raggedcolumns
\setlength{\parindent}{0pt}
\setlength{\parskip}{0pt}
\small
\setlength{\baselineskip}{1.25\baselineskip}

Bin Lin\\
Bo Zhao\\
Boyang Wang\\
Boyang Zhang\\
Boyong Wu\\
Chao Yan\\
Chen Geng\\
Chen Wu\\
Cheng Yi\\
Chengli Feng\\
Chenglin Zhu\\
DanNi Wan\\
Daxin Jiang\\
Dongqing Pang\\
Fei Tian\\
Feng Tian\\
Future Li\\
Gang Yu\\
Guanglong Yang\\
Jia Peng\\
Jiahao Song\\
Jiamin Fan\\
Jiangjie Zhen\\
Jianzheng Gao\\
Jun Chen\\
Li Xie\\
Lifang Zhang\\
Lingli Ji\\
Liying Shi\\
Lun Cai\\
Min Xu\\
Na Wang\\
Peilin Li\\
Peng Yang\\
Pengfei Tan\\
Qingjian Lin\\
Ruijie Xiong\\
Runze Li\\
Shenghua Hu\\
Shi Qiu\\
Siqi Tu\\
Siyi Zhou\\
Tianjiao Deng\\
Wanying Lu\\
Weiming Niu\\
Wen Sun\\
WenWen Qu\\
Xiangyu Zhang\\
Xianwei Zhang\\
XiaoSu Su\\
Xing Chen\\
Xinyu Liu\\
Xuerui Yang\\
Yang Li\\
Yang Yang\\
Yechang Huang\\
Yibo Zhu\\
Yifan Zhang\\
Yiyang Xu\\
Yu Fu\\
Yu Luo\\
Yu Zhou\\
Yumang Wang\\
Yunzhou Ju\\
Yuxiang Yang\\
Zekai Liu\\
Zengwei Yao\\
Zhenwei Mou\\
Zheqi Dai\\
Zhiyue Wu\\
Zichao Zhou

\end{multicols}
\clearpage

\setlength{\bibsep}{0.5\baselineskip}

\bibliography{references}

\end{document}